\documentclass[twocolumn,pre]{revtex4}

\usepackage{epsfig,graphicx}
\newcommand{\nn}{\nonumber}
\newcommand{\eps}{\varepsilon}

\usepackage{times,mathptmx}
\usepackage[T1]{fontenc}
\usepackage{textcomp} 

\begin{document}

\title{Block Copolymers in Electric Fields: A Comparison of Single-Mode
and Self-Consistent Field Approximations}
\author{Yoav Tsori}
\affiliation{Department of Chemical Engineering, Ben Gurion
University, Beer Sheva 84105, Israel. [tsori@bgu.ac.il]}
\author{David Andelman}
\affiliation{ School of Physics and Astronomy, Raymond and Beverly
Sackler Faculty of Exact Sciences, Tel Aviv University, Ramat Aviv
69978, Tel Aviv, Israel. [andelman@post.tau.ac.il]}
\author{Chin-Yet Lin and M. Schick}
\affiliation{Physics Department, Box 351560, University of
Washington, Seattle, WA 98195, USA. [chimney@u.washington.edu];
[schick@mahler.phys.washington.edu]}


\date{August 2, 2005} 

\begin{abstract}

We compare two theoretical approaches to dielectric diblock
copolymer melts in an external electric field. The first is a
relatively simple analytic expansion in the relative copolymer
concentration, and includes the full electrostatic contribution
consistent with that expansion.  It is valid close to the
order-disorder transition point, the weak segregation limit. The
second employs self-consistent field (SCF) theory and includes the
full electrostatic contribution to the free energy at any copolymer
segregation. It is more accurate but computationally more intensive.
Motivated by recent experiments, we explore a section of the phase
diagram in the three-dimensional parameter space of the block
architecture, the interaction parameter and the external electric
field. The relative stability of the lamellar, hexagonal and
distorted body-centered-cubic (bcc) phases is compared within the
two models. As function of an increasing electric field, the
distorted bcc region in the phase diagram shrinks and disappears
above a triple point, at which the lamellar, hexagonal and distorted
bcc phases coexist. We examine the deformation of the bcc phase
under the influence of the external field. While the elongation of
the spheres  is larger in the one-mode expansion than that predicted
by the full SCF theory, the general features of the schemes are in
satisfactory agreement. This indicates the general utility of the
simple theory for exploratory calculations.
\end{abstract}

\maketitle

\section{Introduction}

Block copolymers (BCP) consist of several chemically
distinct sub-chains. They are not only interesting as a model system for
self-assembly, but also for  their chemical versatility and
affordability which have enabled their use in applications such as photonic
waveguides \cite{fink}, tough plastics \cite{leibler-nmat}, ordered
arrays of nano-wires \cite{russell1}, etc. At a given chemical
architecture and temperature, there is one thermodynamically stable
meso-phase, with typical length-scales comparable to the chain size
($\sim10$-$500$ nm). However, the material is rarely perfectly
ordered, but rather is composed of many randomly oriented grains of
size $\sim 1$ $\mu$m. This has an adverse effect on
nanotechnological applications.

A useful way to achieve improved long-range order is to subject the
BCP sample above its glass transition to an external electric field
${\bf E}_0$. Due to the coupling between the field and the
spatially-varying dielectric constant $\kappa({\bf r})$, there is a
preferred orientation of the grains  with respect to the field
\cite{AH93,AH94,PW99,russell2,muthu,onuki95,TA02,TATL-prl,TTL-mm,zvelin}.
It has been shown by Amundson {\it et al.} \cite{AH93,AH94} that the
electrostatic free-energy penalty associated with dielectric
interfaces which are not parallel to the electric field
direction is the driving force for  structures to reorient
so that their interfaces are parallel to the field ($\nabla
\kappa({\bf r})$ perpendicular to ${\bf E}_0$).
 While the free energy penalty can
be eliminated by this reorientation of lamellae and cylinders, it
cannot be eliminated in the body-centered-cubic (bcc) phase, but
only reduced by distorting the bcc spheres.  Thus, the free energy
of this distorted bcc phase, whose symmetry is reduced to
$R{\bar3}m$, increases with respect to the full disordered liquid
(dis), lamellar (lam) and hexagonal (hex) phases \cite{TATL-prl}, a
circumstance which can bring about a phase transition. The effect of
the electric field on the BCP morphology has been substantially
accounted for recently \cite{EAS} by incorporating the electrostatic
Maxwell equations in the full set of self-consistent field (SCF)
equations, which permits calculation of the phase diagram at
arbitrary degrees of segregation.

In this paper we compare two
theoretical  approaches to such a system; the aforementioned SCF study and a
simple analytical approximation consisting of a Ginzburg-Landau
expansion of the free-energy \cite{leibler80}, valid only close to
the order-disorder temperature (ODT).
The paper is organized as follows. In Section 2 we present the
free-energy model which includes the electrostatic energy of the
BCP in the field. In Section 3 we calculate the way in which an
initial mesophile deforms under the influence of the field, and
also find the relative stability of the competing phases. A
comparison is made with the results of the SCF model. Section 4
contains a brief conclusion.

\section{Model}

Although the effect we consider here is generic to any multi-block
BCP melts, we will restrict the discussion in this paper to the
simplest A/B {\it di}-block copolymer, where a spatial variation of
the relative A/B monomer concentration yields a spatial dependence
of the dielectric constant and, hence, of the response to an
external electric field.

We also assume for simplicity that the A monomeric volume is equal
to the B one. Then the volume fraction of the A monomers $f$, ($0\le
f\le 1$), is equal to its molar fraction. The order parameter
$\phi({\bf r})$ is defined as the local deviation of the A-monomer
concentration $\phi_A({\bf r})$ from its average value: $\phi({\bf
r})=\phi_A({\bf r})-f$. From an incompressibility condition of the
melt we also have at each point ${\bf r}$,  $\phi_B({\bf
r})=1-\phi_A({\bf r})$. In the absence of any external electric
fields, the bulk BCP free-energy per polymer chain, $F_b$, in units
of $k_BT$, can be written as a functional of the order parameter,
$\phi({\bf r})$. One way of generating a simple analytical expansion
in the order parameter relies on a Ginzburg-Landau-like free energy,
which can be justified close to the order-disorder point (ODT)
\cite{leibler80,OK86,FH87} and is repeated here without further
justifications:

\begin{eqnarray}\label{Fb}
F_b=\frac{1}{\Omega}\int\left\{\frac12\tau\phi^2+\frac12
h\left(\nabla^2\phi+q_0^2\phi\right)^2+\frac{\lambda}{6}\phi^3+\frac{u}
{24}\phi^4\right\}{\rm d}^3 r\nn\\
\end{eqnarray}
where  $\Omega$ is the system volume, and
\begin{eqnarray}
\tau&=&\frac{2N(\chi_s-\chi)}{c^2}\nn\\
q_0\sim 1/R_g&=&\left(\frac16 Nb^2\right)^{-1/2}\nn\\
h&=&\frac32 \frac{R_g^2}{q_0^2}
\end{eqnarray}
$b$ is the Kuhn length, $R_g$ the radius of gyration, $\chi$ is the
Flory parameter, $N=N_A+N_B$  the total chain polymerization index,
$N\chi_s$ the spinodal value of $\chi N$ \cite{leibler80}, $c$ is a
constant of order $1$, and $\lambda$ and $u$ are functions of $f$ as
in refs. \cite{leibler80,OK86,FH87}. The phase diagram in the
($f$,$\chi N$) plane, as derived from the free-energy, eq~\ref{Fb},
is symmetric with respect to exchange of $f$ and $1-f$. For small
values of $\chi\sim 1/T$, the melt is disordered: $\phi({\bf r})=f$
is constant. For $\chi N$ larger than the ODT value of $\simeq 10.5$
and for nearly symmetric BCP composition ($f\approx \frac12$), the
lamellar phase is the most stable. As $|f-\frac12|$ increases, the
stable phases are doubly-connected gyroid, hexagonal and bcc phases
\cite{leibler80,OK86,MS-prl}.

Let us now consider a BCP slab placed in an external electric field,
${\bf E_0}$. The free-energy per polymer chain, again in units of
$k_BT$, is $F_{\rm tot}=F_b+F_{\rm es}$, where the electrostatic
energy contribution $F_{\rm es}$ is given by the integral
\begin{eqnarray}\label{Fes}
F_{\rm es}=-\frac12\frac{\eps_0v_p}{k_BT\Omega}\int\kappa({\bf
r})\left[\nabla\psi({\bf r})\right]^2~{\rm d}^3r
\end{eqnarray}
Here $\eps_0$ is the vacuum permittivity, $\kappa({\bf r})$ is the
local dielectric constant, $v_p$ is the volume per chain, and $\psi$
is the electrostatic potential obeying the proper boundary
conditions on the electrodes. The local field is ${\bf
E(r)}=-\nabla\psi$. We note that the variation of $F_{\rm es}$ with
respect to $\psi$ yields
\begin{eqnarray}
\nabla\cdot\left(\eps_0\kappa({\bf r})\nabla\psi\right)=0
\label{maxwell}
\end{eqnarray}
which is the usual Maxwell equation ${\nabla\cdot} {\bf D}=0$ for
the displacement field ${\bf D}=\eps_0\kappa{\bf E}$. We consider
a simple geometry of a BCP slab filling the gap between two parallel
and flat electrodes separated by a distance $d$ and potential
difference $V$. Even when a non-homogeneous dielectric material like
a BCP fills the gap between the two electrodes, the spatially
averaged electric field in between the electrodes $\langle {\bf
E}\rangle$ is constrained to be $E_0=V/d$. The local field ${\bf
E(r)}$ differs from its average  due to the nonuniformity of the
dielectric constant, since $\kappa=\kappa(\phi)$ depends on the local
concentration $\phi({\bf r})$ through a constitutive equation. In
this paper we assume for simplicity a linear constitutive relation,
\begin{eqnarray}
\kappa({\bf r})&=&\langle\kappa\rangle+\phi({\bf
r})\Delta\kappa,~~~~~~~~\Delta\kappa\equiv\kappa_A-
\kappa_B,\nn\\
\langle\kappa\rangle&=&f\kappa_A+(1-f)\kappa_B,
\label{constit}
\end{eqnarray}
where throughout this paper we use $\kappa_A=6.0$ and
$\kappa_B=2.5$, thus modelling an A/B diblock copolymer where the A block is
polymethylmethacrylate (PMMA) and the B block is polystyrene (PS),
as is used in many experiments. Other constitutive relations can be considered
\cite{foot}.

When a field is applied on a melt in the lamellar or hexagonal
phases, it exerts torque which causes sample rotation. The torque is
zero, and the energy lowest, when the lamellae or cylinders are oriented parallel to the
field. In such states, as well as the disordered phase, the electrostatic energy, eq~\ref{Fes}, of
the system is equal to a reference energy, given in eq \ref{reference}.
The bcc array of spheres, on the other hand, always
has dielectric interfaces that are not  parallel to the field,
and its electrostatic energy is higher than the reference value.
Hence, the spheres elongate in
the applied field direction, to an extent which is a balance between
electrostatic and elastic forces, as calculated below.

The reference energy per polymer chain, in units of $k_BT$, is
simply
$-\frac12\langle \kappa \rangle {\hat E}_0^2$, where ${\hat E_0}$ is
the applied field measured in the natural unit
$(k_BT/\eps_0v_p)^{1/2}$,
\begin{eqnarray}
\hat{E}_0=\left(\frac{\eps_0 v_p}{k_BT}\right)^{1/2}E_0
\label{reference}
\end{eqnarray}

 Let us estimate the value of the actual applied field
corresponding to $\hat{E}_0=1$. At $100^\circ$\,C and using typical
polymer volume per chain in the range $v_p\simeq 50 - 250$ nm$^3$,
we find $E_0\simeq 47-107$ V/$\mu$m. This is a relatively large
field that can cause dielectric breakdown in some BCP films.
Therefore, the experimentally interesting regime is usually
$\hat{E}_0\lesssim 1$.

The free-energy $F_{\rm tot}$ as formulated above is valid close to
the ODT point (weak segregation limit), where the concentration
variations are small, $\phi({\bf r})\ll 1$, and therefore the
analysis can be carried out within the so-called {\it one-mode
approximation}. Motivated by recent experiments
\cite{TATL-prl,zvelin}, we concentrate on the transition from
distorted spheres to cylinders or disordered melt in presence of an
applied electric field. Taking ${\bf E}_0$ to be in the
($1$,$1$,$1$) direction, we write the order parameter $\phi$  as a
linear superposition of six components
\begin{equation}
\phi=w\phi_1+g\phi_2 \label{def_gw}
\end{equation}
where
\begin{eqnarray}
 \phi_1&=&\sum_{i=1}^3\cos({\bf q}_i\cdot{\bf r}),~~~~~~~~
\phi_2=\sum_{i=1}^3\cos({\bf k}_i\cdot{\bf r})\label{def_f23}
\end{eqnarray}
The $q$'s and $k$'s are wave-vectors given by
\begin{eqnarray}
{\bf q}_1&=&\frac{q_0}{\sqrt{2}}(-1,0,1),~~~~~~~{\bf k}_1=\frac{q_0}{\sqrt{2}}(1,0,1),\nn\\
{\bf
q}_2&=&\frac{q_0}{\sqrt{2}} (1,-1,0),~~~~~~~{\bf
k}_2=\frac{q_0}{\sqrt{2}} (1,1,0),\nn\\
{\bf
q}_3&=&\frac{q_0}{\sqrt{2}}(0,1,-1)~~~~~~~~ {\bf
k}_3=\frac{q_0}{\sqrt{2}}(0,1,1)\label{def_kq}
\end{eqnarray}
and all have the same magnitude $q_0$. The three linearly dependent
${\bf q}_i$ are orthogonal to the ($1$,$1$,$1$) direction and
describe a hexagonal phase with axis along that direction. The three
${\bf k}_i$ have equal and non-zero projections on the ($1$,$1$,$1$)
axis. The six wavevectors transform into one another under the
symmetry operations of the bcc phase. In the absence of an external
field, each of these wavevectors would contribute equally in the
order parameter expansion, so that $g$ and $w$ would be equal. These
wavevectors characterize the first mode in such an expansion. Hence
the name of the approximation.

The amplitudes $w(E_0)$ and $g(E_0)$ depend on the magnitude of the
average external field $E_0$. Depending on the values of the two
amplitudes, $g$ and $w$, we can represent the order parameter of all
phases of interest in the form of eq~\ref{def_gw}: $w=g\ne 0$
represents an undistorted bcc, while an $R{\bar 3}m$ (distorted bcc)
phase  oriented along the ($1$,$1$,$1$) direction is represented by
two non-zero amplitudes $w\neq g$. A hexagonal phase of  cylinders
whose long axis is in the $(1,1,1)$ direction has  $w\ne 0$ and
$g=0$. And finally, $g=w=0$ represents the disordered melt. As was
mentioned above, the spatially averaged electric field is simply the
magnitude of the external field, $E_0$. However, local changes in
$\phi({\bf r})$ give rise to local changes in $\kappa({\bf r})$.
Consequently, the electric field can be written as follows:
\begin{eqnarray}
{\bf E}&=&{\bf E}_0+\delta{\bf E}
\end{eqnarray}
where $\delta{\bf E}$ is the deviation from the average. The symmetry of
the electrostatic potential follows from the Maxwell equation, eq.
\ref{maxwell}, the constitutive equation, eq. \ref{constit}, relating
the dielectric constant to the order parameter, and the $R{\bar3}m$
symmetry of the distorted bcc phase. From the symmetry of the potential,
one finds that $\delta
{\bf E}$ can be decomposed in terms of the same bases of $k$'s and $q$'s
given above:

\begin{eqnarray}
\delta {\bf E}&=&
\alpha{\bf E}_1+\beta{\bf E}_2\nonumber\\
{\bf E}_1&=&E_0\sum_{i=1}^3\hat{\bf q}_i\cos({\bf q}_i\cdot{\bf
r}),\nn\\
{\bf E}_2&=&E_0\sum_{i=1}^3\hat{\bf k}_i\cos({\bf
k}_i\cdot{\bf r})\label{e23}
\end{eqnarray}
where ${\bf \hat{k}}_i={\bf k}_i/|{\bf k}_i|$ and ${\bf
\hat{q}}_i={\bf q}_i/|{\bf q}_i|$ are unit vectors. Within the
one-mode approximation, the local field at each point is given in
terms of two amplitudes, $\alpha$ and $\beta$. They determine the
projection of the deviation field $\delta{\bf E}$  onto the ${\bf
E}_1$ or ${\bf E}_2$ directions.

In order to proceed we insert the $E$-field expressions,
eq~\ref{e23}, into the electrostatic free-energy, eq~\ref{Fes}.
Using the definitions of eqs~\ref{def_gw}-\ref{def_kq} and the
properties:
\begin{eqnarray}
{\bf q}_i\cdot(1,1,1)=0 ~~;~~ |{\bf q}_i|^2=q_0^2 ~~;~~ {\bf
q}_i\cdot{\bf q}_j=-q_0^2/{2}, ~~{ i\ne j}~~~
\end{eqnarray}
and
\begin{eqnarray}
{\bf k}_i\cdot(1,1,1)=2q_0/\sqrt{2} ~~~;~~ |{\bf k}_i|^2=q_0^2 ~~;~~
{\bf k}_i\cdot{\bf k}_j=q_0^2/2,~~{ i\ne
j}~~~~~\nonumber\\
\sum_{i=1}^3{\bf k}_{i}=\frac{2q_0}{\sqrt{2}}(1,1,1)~~;~~~ {\bf
k}_i\cdot{\bf q}_i=0 ~~;~~~{\bf k}_i\cdot{\bf
q}_j=\pm\frac{q_0^2}{2},~~i\ne j~~~~~
\end{eqnarray}
we can perform the rather straightforward spatial averages of the
various terms in the free energy, eq~\ref{Fes}, and obtain:
\begin{eqnarray}
F_{\rm es}&=&-\frac12 \frac{\hat{E}_0^2}{\Omega}\int\kappa({\bf
r})\left[\frac{{\bf E}({\bf r})}{E_0}\right]^2~{\rm d}^3r
\nonumber\\
&=& -~\frac34\langle \kappa \rangle(\alpha^2+\beta^2){\hat
E}_0^2+\frac38w\Delta\kappa  (\alpha^2-\beta^2){\hat
E}_0^2\nn\\
&-&\sqrt{\frac32}\beta g\Delta\kappa {\hat
E}_0^2-\frac12\langle \kappa \rangle {\hat E}_0^2
\end{eqnarray}
The last term is simply the reference energy which is common to all
phases.

For a given state of $\phi$ (a given BCP morphology), which is
determined by a given value of $w$ and $g$, the values of
$\alpha(w,g)$ and $\beta(w,g)$ are determined by the
Maxwell equation, eq \ref{maxwell}.
This is equivalent to obtaining them by varying
$F_{\rm es}$ with respect to $\alpha$ and $\beta$. The procedure yields
$\alpha=0$ and $\beta=-\sqrt{2/3}\Delta\kappa g/(\langle \kappa
\rangle+\frac12\Delta\kappa w)$. Thus,
 $F_{\rm es}$ is given by
\begin{equation}\label{Delta_Fes}
F_{\rm es}=\left[\frac{(\Delta\kappa)^2}{2\langle \kappa
\rangle+w\Delta\kappa } g^2-\frac12 \langle\kappa\rangle\right]{\hat E}_0^2
\end{equation}
It is instructive to compare this result with the perturbation
expression used by Amundson, Helfand and co-workers \cite{AH93,AH94},
\begin{eqnarray}
F_{\rm es}^{\rm
AH}=\left[\frac{(\Delta\kappa)^2}{2\langle \kappa \rangle}g^2 -\frac12
\langle\kappa\rangle
\right]{\hat E}_0^2,
\label{ad}
\end{eqnarray}
a result  obtained from eq~\ref{Delta_Fes} in the limit
$\Delta\kappa/\langle \kappa \rangle\to 0$. In particular, the
result of eq \ref{ad} yields a free energy which is symmetric
under the interchange of monomers A and B, while the result of
eq~\ref{Delta_Fes} does not.  Such a symmetry is not expected in
general: a system of ultra-high $\kappa$ spheres (metallic limit)
embedded in an insulator matrix has a different energy than the
system of insulating spheres embedded in a ultra-high $\kappa$
matrix, even when the average dielectric constant is the same
$\langle\kappa\rangle$.

Employing the single-mode Ansatz $\phi=w\phi_1+g\phi_2$ in
eq~\ref{Fb}, we finally obtain for the total free-energy per
polymer chain in units of $k_BT$
\begin{eqnarray}
F_{\rm tot}&=&\frac{3\tau}{4}(w^2+g^2)~+~\frac{\lambda}{4} w(3g^2+w^2)
~+~\frac{15}{64}u(g^4+4g^2w^2+w^4)\nn \\ &+& \left[\frac{
(\Delta\kappa)^2}{\left(2\langle \kappa \rangle+w\Delta\kappa
\right)}g^2-\frac12\langle\kappa\rangle\right]\hat{E}_0^2 \label{Ftot}
\end{eqnarray}

In the next section we minimize this energy with respect to $w$ and
$g$ at a given dimensionless external field $\hat{E}_0$ and polymer
architecture $f$, calculate the elongation the spheres of the bcc phase,
and obtain the phase diagram.

\section{Results}
%

\begin{figure}
\centering
\includegraphics[scale=0.55,bb=75 245 495 575,clip]{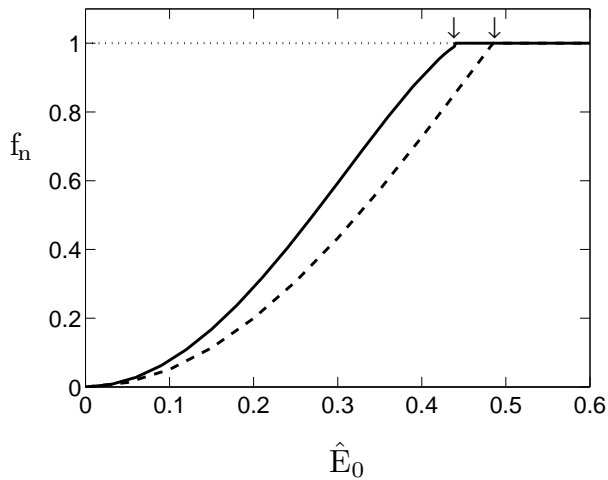}
\caption{Normalized free-energy per polymer chain $f_n$, defined in
eq \ref{normf}, of the distorted bcc
 phase ($R{\bar3}m$) as a function of dimensionless field
$\hat{E}_0$. The system is characterized by $f=0.3$ and $\chi
N=14.4$. We compare the one-mode calculation (solid line) as
obtained from minimization of eq~\ref{Ftot} with a SCF calculation
(dashed line). The $R{\bar3m}$ phase in the SCF calculation has a
lower free energy that the solid line (one mode), and crosses the
hex energy at higher value of $\hat{E}_0$ of about $0.49$, while
the one-mode approximation crosses at $\hat{E}_0 \approx 0.43$
(both marked with arrows). In this figure and following ones we
used $\kappa_A=6$ and $\kappa_B=2.5$, modelling a PMMA-PS
copolymer.}
\end{figure}

As noted above, the functional form $\phi=w\phi_1+g\phi_2$ allows us
to distort smoothly a bcc array of spheres (having nonzero $w=g$)
via a distorted bcc phase ($w\ne g$), and into a hexagonal array of
cylinders (non-zero $w$ but with $g=0$). The disordered phase is
given by $w=g=0$. One is thus able to obtain the full phase diagram
by minimizing eq~\ref{Ftot} with respect to the amplitudes $w$ and
$g$.

\begin{figure}
\centering
\includegraphics[scale=0.6,bb=105 220 490 575,clip]{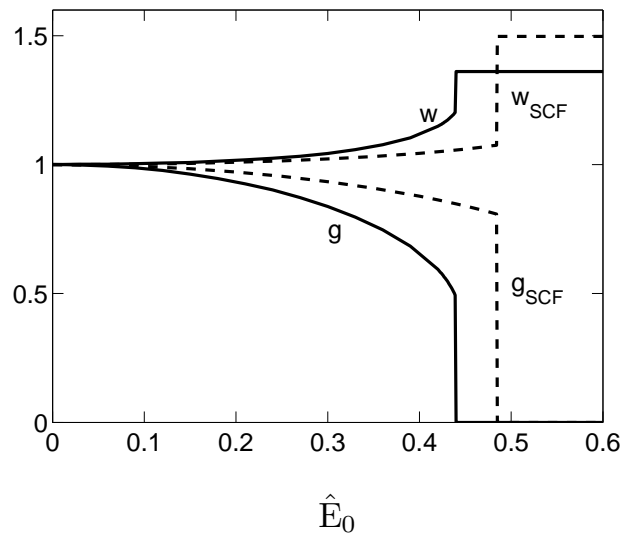}
\caption{The amplitudes $w$ and $g$ normalized by their common value
at zero $E$-field, $g(0)=w(0)$, as a function of dimensionless
external field $\hat{E}_0$. Solid line: one-mode approximation. The
amplitudes have a discontinuous jump  at $\hat{E }_0\simeq 0.43$,
where the structure contains cylinders oriented along the field
($g=0$). Dashed lines: the same, but taken from a multi-mode SCF
calculation with a jump at higher $E$ values of $\hat{E }_0\simeq
0.49$. All parameters as in figure 1.}
\end{figure}
Before presenting the phase diagram, let us consider a point in
the $(f,N\chi)$ plane for which the stable phase at zero $E$ field
has a bcc symmetry. For presentation purposes, in figure~1 we have
subtracted from the free energy the reference electrostatic
energy, $-\langle\kappa\rangle{\hat E}_0^2/2$, common to all
phases, also subtracted the total free energy of the bcc phase in
zero field, $F_{\rm tot}^{\rm bcc}(0)$, and normalized the
resulting free energy by that of the hex phase in zero field; that
is we have plotted
\begin{equation}
f_n\equiv \frac{F_{\rm tot}({\hat E}_0)+\langle\kappa\rangle{\hat
E}_0^2/2-F_{\rm tot}^{\rm bcc}(0)}{F_{\rm tot}^{\rm hex}(0)}
\label{normf}
\end{equation}

In the figure we show how the free energy $f_n$ changes with
$\hat{E}_0$ for $f=0.3$ and $N\chi=14.4$.
At $\hat{E}_0=0$ the bcc is the stable phase, and its free energy
increases with increasing field $\hat{E}_0$, until it equals the
free energy of the hex phase at a transition field $\hat{E}_0\simeq
0.43$.  At larger fields the stable structure is a hex phase of
cylinders oriented along the external field ${\bf E}_0$.
The solid line in figure 1 is the result obtained from the
one-mode approximation given above, while the dashed line is a
obtained from the SCF theory,
(as in Ref~\cite{EAS}). It has a lower free-energy. Consequently,
the transition field in the SCF framework is higher and occurs at
about $\hat{E}_0\simeq 0.49$.

Figure 2 is a plot of the amplitudes $w(\hat{E}_0)$ and
$g(\hat{E}_0)$, normalized by their zero-field value
$w(\hat{E}_0=0)=g(\hat{E}_0=0)$. Both amplitudes start at their
common value in the undistorted bcc phase. As the field increases,
$w$ increases while $g$ decreases. The spheres  elongate in the
direction of the field as a result of competition between
electrostatic and elastic forces. At the transition field, there is
a sharp, discontinuous, transition in the order parameter. Above
this field, $w$ attains a fixed value while $g$ drops abruptly to
zero. In this state the BCP morphology is that of cylinders oriented
parallel to the external field. The dashed lines correspond to the
values obtained from the SCF theory. Clearly, in the one-mode approximation,
the spheres' deformation and eccentricity are larger than in the
SCF theory.
\begin{figure}
\centering
\includegraphics[scale=0.45,bb=15 180 545 590,clip]{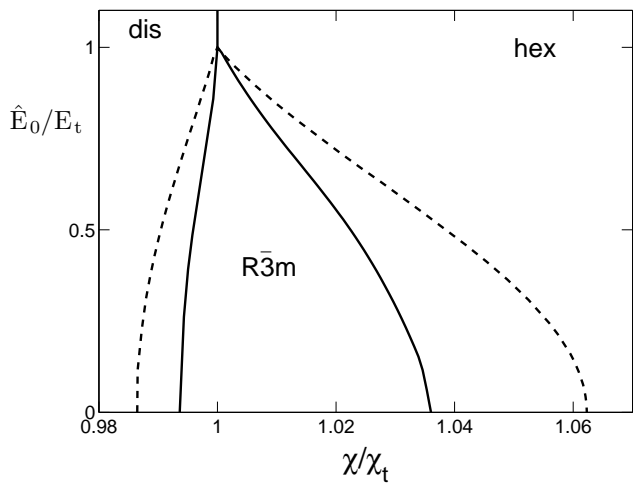}
\caption{Phase diagram in the ($\chi/\chi_t$, $\hat{E}_0/E_t)$
rescaled plane for a fixed value of BCP asymmetry, $f=0.3$. The
distorted bcc ($R{\bar3}m$) region is confined between two
transition lines which terminate at a triple point $(\chi_t,E_t)$.
The other two phases are the disordered one (dis) and the
hexagonal one (hex). The solid lines are the prediction of the
one-mode approximation, with axes scaled by the appropriate
$\chi_t\simeq 14.3/N$ and $E_t\simeq 0.49$. Dashed lines are the
SCF calculation scaled by the appropriate SCF values of
$\chi_t\simeq 14.11/N$ and $E_t\simeq 0.668$. }
\end{figure}

The above calculation can be repeated for any $(f,N\chi)$  and
$\hat{E}_0$ field values and allows the construction of  the full
three-dimensional phase diagram in the $(f,N\chi,\hat{E}_0)$
parameter space. In figure 3 we present a cut of the phase diagram
at fixed $f=0.3$. The region of a stable $R{\bar3}m$ phase
(distorted bcc) is bound by two lines of phase transitions: one
between this phase and the disordered phase, and the other between
it and the hex phase. These two lines meet at the triple point
$(\chi_t,E_t)$. In figure 3, the different triple point values
obtained from the two calculations are used to rescale both axes:
$\chi/\chi_t$ and $\hat{E}_0/E_t$. At fields larger than $E_t$ the
$R{\bar3}m$ is not stable at any value of $\chi$. The solid lines in
the figure are the one-mode prediction, while the dashed lines are
obtained with the SCF calculation. The values of $E_t$ are $0.49$
and $0.67$ for the two theories, respectively.

For an additional comparison between the theories, we have
examined the case in which, at a fixed value of $f=0.3$, the
dielectric constants of the majority and minority components are
interchanged ({\em i.e.} $\kappa_A=6.0 \leftrightarrow
\kappa_B=2.5$, hence, $\Delta\kappa \rightarrow -\Delta\kappa$).
In both theories we find an increase in the value of the external
field needed to bring about a transition from the distorted bcc
phase to the hex phase. Thus, this subtle effect, which is not
captured by the perturbation result of eq~\ref{ad}, is obtained in
the simple one-mode approximation, eq~\ref{Delta_Fes}.

\section{Conclusions}

A simple theory for  a non-homogeneous diblock copolymer (BCP) melt
in an external electric field is presented, and compared with a more
accurate, but more computationally intensive self-consistent field
(SCF) one. The differences between the two theories in {\em zero}
external field are well known. In particular, the accuracy of the
phase boundaries produced by the one-mode approximation deteriorates
outside the vicinity of the ODT point (weak segregation) as compared
to the  SCF theory \cite{MS-prl}.  However, as in the zero field
case, the qualitative behavior of the system in the presence of a
field is described surprisingly well. The full electrostatic
free-energy contribution is used, consistent within the one-mode
approximation, eq~\ref{Delta_Fes}. This was not accounted for in
previous analytical studies~\cite{AH93,AH94,TATL-prl,zvelin}, where
only quadratic terms in the electrostatic potential were retained.
The simple one-mode approximation captures the elongation of the
spheres of a bcc phase when placed under an external field. The
elongation is  in the direction of the applied $E_0$ field. The two
amplitudes describing this elongation, $w$ and $g$, as shown in
figure 2. At a threshold value of the electric field, a first-order
transition to a hexagonal phase occurs and the amplitudes jump
discontinuously.

As shown in figure 3, the simple, analytic, one-mode approximation also captures
the essence of the phase diagram; the reduction in the phase space
occupied by the distorted bcc phase as the field increases, and its
eventual disappearance at a triple point.

Lastly the one-mode theory also captures the subtle interplay between
structure and electrostatic response as evidenced by its prediction of
a different critical field for phase transitions when the dielectric constants
of the constituents are interchanged, a prediction in agreement
with the more accurate theory \cite {EAS}.

Given its ability to capture all of the above effects, and given its extreme simplicity,
such a theory could serve for useful exploratory studies in other problems concerning the
effect of electric fields on block copolymers.

\section*{Acknowledgements}

We have benefited from discussions with L. Leibler, T. Russell and T. Xu.
Support from the U.S.-Israel Binational Science Foundation (B.S.F.)
under grant No. 287/02,  the Israel Science Foundation under
grant No. 210/01, and the National Science Foundation under
Grant No. 0140500  is gratefully acknowledged.


\begin{thebibliography}{99}
\bibitem{foot}
We briefly mention that a quadratic constitutive relation between $\eps$
and $\phi$, $\eps=\eps^{(0)}+\eps^{(1)}\phi+\eps^{(2)}\phi^2$, was
previously investigated in refs~\cite{AH93,AH94,nature}. The
existence of the quadratic term  leads to an upwards shift in the
ODT temperature:
%
$\Delta T_{\rm ODT}={v_p{\eps}^{(2)}}/{k_B}E_0^2$,
%
where $v_p=\Omega/n$ is the volume of one BCP chain. This shift is
typically small.
\end{thebibliography}
\end{document}